\documentclass[journal]{vgtc}                     

\onlineid{0}

\vgtccategory{Research}

\vgtcpapertype{please specify}

\title{Projection Mapping under Environmental Lighting\\ by Replacing Room Lights with Heterogeneous Projectors}

\author{%
  \authororcid{Masaki Takeuchi}{0000-0002-0773-7705},
  \authororcid{Hiroki Kusuyama}{0009-0007-9406-2120},
  \authororcid{Daisuke Iwai}{0000-0002-3493-5635}, and
  \authororcid{Kosuke Sato}{0000-0003-1429-9990}
}

\authorfooter{

   \item 
        Masaki Takeuchi, Hiroki Kusuyama, Daisuke Iwai, and Kosuke Sato are with Graduate School of Engineering Science, Osaka University. 
        E-mail: see https://www.sens.sys.es.osaka-u.ac.jp.
}

\abstract{
Projection mapping (PM) is a technique that enhances the appearance of real-world surfaces using projected images, enabling multiple people to view augmentations simultaneously, thereby facilitating communication and collaboration.
However, PM typically requires a dark environment to achieve high-quality projections, limiting its practicality.
In this paper, we overcome this limitation by replacing conventional room lighting with heterogeneous projectors.
These projectors replicate environmental lighting by selectively illuminating the scene, excluding the projection target.
Our contributions include a distributed projector optimization framework designed to effectively replicate environmental lighting and the incorporation of a large-aperture projector, in addition to standard projectors, to reduce high-luminance emitted rays and hard shadows---undesirable factors for collaborative tasks in PM.
We conducted a series of quantitative and qualitative experiments, including user studies, to validate our approach.
Our findings demonstrate that our projector-based lighting system significantly enhances the contrast and realism of PM results even under environmental lighting compared to typical lights.
Furthermore, our method facilitates a substantial shift in the perceived color mode from the undesirable aperture-color mode, where observers perceive the projected object as self-luminous, to the surface-color mode in PM.
} 


\teaser{
  \centering
  \includegraphics[width=\linewidth, alt={Projection Mapping under Environmental Lighting}]{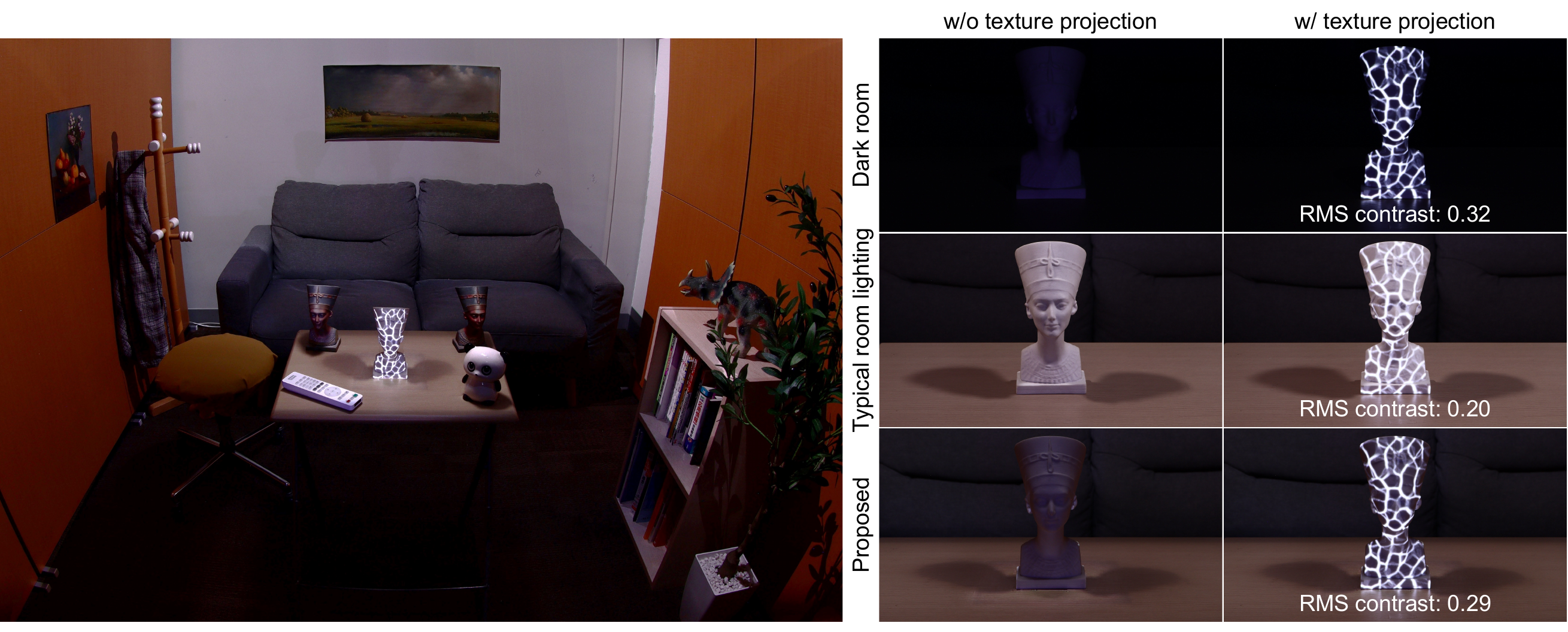}
  \caption{The proposed projection mapping system mitigates the reduction in contrast under environmental lighting. The left image provides an overall view of the texture projection in a well-lit room using the proposed technique. The crack pattern texture is projected onto the white statue located at the center of the desk. The images on the right show enlarged views of the statue without (left) and with (right) the texture projection. These images were captured in three different lighting conditions: a dark room, where projection mapping is typically conducted; under typical room lighting with LED lights; and under the proposed environmental lighting using projectors. The results demonstrate that our technique reproduces typical room lighting using multiple projectors and enhances the contrast of the projected result on the target surface.}
  \label{fig:teaser}
  }

\usepackage{tabu}                      
\usepackage{booktabs}                  
\usepackage{lipsum}                    
\usepackage{mwe}                       

\usepackage{mathptmx}                  

\usepackage{comment}
\usepackage{amsmath}
\usepackage{amssymb}
\usepackage{siunitx}
\usepackage{here}
\usepackage{url}

\begin{document}


\firstsection{Introduction}

\maketitle


Projection mapping (PM) is a technology that allows computer-generated images to be superimposed on a target object in physical space using projectors. 
Previous studies have investigated the utility of PM in a range of fields, including, but not limited to, medicine~\cite{00000658-201806000-00024}, teleconferencing~\cite{10.1145/2818048.2819965,8172039,10.1145/280814.280861}, museum guides~\cite{SCHMIDT20191,1377099}, makeup~\cite{https://doi.org/10.1111/cgf.13128,8007312}, object searches~\cite{10.1145/1015706.1015738,10.1145/1180495.1180519,10.1145/1959826.1959828,8007248,Iwai2011}, urban planning~\cite{10.1145/302979.303114}, and artwork creation~\cite{10.1145/2366145.2366176,10.1145/1166253.1166290,970539}.
However, there is a strong and crucial constraint in PM---images need to be projected in dark environments.
Typical environment lighting globally elevates the scene brightness.
Therefore, if PM is performed under it, the dark areas of a projected result inevitably get brighter, which causes significant contrast degradation in the projected result.
This constraint has become so ingrained that it is unquestionably accepted as the norm.
However, it significantly limits the application fields of PM because PM in dark environments suffer from various drawbacks.

First, a dark environment makes it difficult for users to see each other's faces.
An important advantage of PM over other types of augmented reality (AR) displays, such as optical see-through and video see-through systems, is that PM allows multiple co-located users to view augmented objects without the need for wearable or handheld devices, facilitating face-to-face discussions about the objects.
However, this advantage is significantly diminished by this constraint.
Secondly, while most PM applications, such as product design~\cite{6949562,CASCINI2020103308,6226394,PARK201538,8797923}, are developed with the goal of visually altering surface materials through projected imagery, the augmentations in these applications appear physically and perceptually unnatural.
Objects are only visible when illuminated by light if they are not self-luminous.
Consequently, it is physically unnatural that only the target object in PM is visible in a dark environment. In particular, PM results must appear completely dark when the altered material is specular or transparent because surrounding environments are reflected and refracted by the specular and transparent objects, respectively.
However, this aspect was not explicitly taken into account in previous work~\cite{10.1145/2816795.2818111,10.1145/3388534.3407297,9714121}.
In terms of the perceptual unnaturalness, despite the absence of explicit investigation into this aspect, it is likely that projected surfaces are perceived in the aperture-color mode (as self-luminous) rather than in the surface-color mode (as illuminated).
Our visual system estimates the highest luminance achievable through reflected light under a given illuminant, and an object appears self-luminous when its luminance exceeds the estimated highest luminance~\cite{10.1167/jov.21.13.3}.
If there is no environmental lighting, the highest luminance should be estimated as zero.
Consequently, any projected results are theoretically perceived in the aperture-color mode, even when the altered material is not inherently self-luminous.

In this study, we propose substituting multiple projectors for typical room lights to selectively illuminate the scene, while excluding the projection target, as a means of overcoming the dark environment constraint of PM.
Projector-based environmental lighting allows for local control of the illuminance and chromaticity incident on scene surfaces, enabling us to avoid undesirable global illuminance elevation on the target surface.
Consequently, we can mitigate the contrast reduction caused by typical environment lighting.
There are two primary contributions aimed at addressing technical challenges unique to the proposed framework.
First, we introduce a novel distributed optimization method for multiple projectors that determines the projector pixel values to accurately reproduce the illuminance and chromaticity incident on surfaces under typical luminaires.
Leveraging the spatially low-frequency nature of environmental lighting, this technique computes pixel values more efficiently than a conventional pixel-wise radiometric optimization framework~\cite{7265057}.
Furthermore, this method takes into account the inter-reflection of projected light between adjacent surfaces.
Inter-reflection significantly increases the illuminance and can lead to inaccurate lighting reproduction results without proper consideration.
Secondly, we propose the use of a custom projector with a much larger aperture than existing off-the-shelf projectors to illuminate the environmental surfaces surrounding the projection target.
As a result, our multi-projection system employs heterogeneous projectors.
Typical luminaires function as area light sources, emitting diffuse light, which results in low-luminance emitted rays and soft shadows when they are occluded.
In contrast, projectors function as point light sources and emit directional light, resulting in high-luminance rays and hard shadows, which could potentially disrupt users in completing their tasks using PM.
To resolve this inconsistency, we utilize the large-aperture projector, which can act as an area light source, to illuminate scene surfaces around the projection target, assuming that multiple observers frequently occlude the light from this projector.

We validate the feasibility of our proposal by installing a prototype in a room and conducting a series of quantitative and qualitative experiments.
First, we demonstrate that our technique can replicate various environmental lighting effects, similar to typical luminaires, using multiple projectors.
Then, in a PM experiment, we illustrate how the system enhances the contrast of projected results on a target surface, thus increasing the realism of PM.
Additionally, we show that the surface material of the target object is naturally transformed into a polished mirror, reflecting the surrounding environments lit by our PM-based lighting.
Finally, we conduct user studies to confirm that PM results presented by our system are perceived more in surface-color mode compared to those under a typical dark room condition, and that our large-aperture projector provides lighting that is more consistent with typical lighting than a standard projector, thus offering better collaborative PM environments.

Our primary contributions are as follows:

\begin{itemize}
    \item Introducing a novel environmental lighting framework that replaces typical room lights with a combination of heterogeneous projectors, including multiple standard projectors and a large-aperture projector, to address the dark environment constraint in PM
    \item Developing a distributed optimization method for the projectors to accurately replicate the environmental lighting of typical luminaires
    \item Employing a large-aperture projector as an area light source to illuminate the surrounding environmental surfaces of the projection target
    \item Demonstrating the feasibility of the proposed methods by installing a prototype in a room
    \item Confirming, through user studies, that observers perceive that the PM in our system alters the reflectance properties of the target surface as if under typical luminaires
\end{itemize}

\section{Related Work}
\label{sec:related_work}

Researchers have worked on covering an entire room by PM, while turning off the room lights.
Jones et al., using multiple projectors, transformed an ordinary room by projecting images onto all of its surfaces, creating a CAVE-like immersive gaming space~\cite{10.1145/2642918.2647383}.
The same research group has also developed a technology that converts an entire meeting room into an information display using PM~\cite{10.1145/3132272.3134117}.
These studies aim to provide immersive experiences without the need for dedicated immersive environments by seamlessly blending projected images from different projectors onto various surfaces within an ordinary environment.
In contrast, our goal is to faithfully reproduce the illuminance and chromaticity incident on the room surfaces under typical room lighting using a multi-projection system.
Therefore, while blending technology plays a crucial role in our system, we also need to estimate the illuminance and chromaticity of room lighting and generate projection images that accurately replicate it.
The PM-based illumination was also explored for occlusion-capable optical see-through AR, realizing augmented imagery to appear bright and opaque even in a lit room~\cite{1115088,6549352}. They also focused on selectively darkening a part of a scene, although they did not address the dark environment constraint in PM. Specifically, accurate room lighting reproduction was still an unsolved issue.

%
%
The estimation of illuminance and chromaticity for a given illuminant has been extensively researched in the fields of computer vision and virtual production~\cite{Abdelhamed_2021_CVPR,10.1145/3543664.3543681,10.1117/12.2071915,10.2312:sr.20221150,10.1145/2897824.2925934}.
Previous techniques employ multiple color patches with known spectral reflectance properties and analyze these captured patches for estimation purposes.
The estimated results are subsequently employed to ensure color constancy in various computer vision tasks and to relight physical objects situated within a virtual production setup, such as a light stage.
While our task shares similarities with the latter scenario, we must additionally account for inter-reflections among the room surfaces, which have a more pronounced impact on the illuminance and chromaticity incident upon the surfaces in our setup compared to a standard virtual production environment.

%
%
A well-studied technology related to illuminant reproduction in PM is radiometric compensation, which is developed to display desired colors on textured surfaces~\cite{1381255,https://doi.org/10.1111/j.1467-8659.2008.01175.x,https://doi.org/10.1111/cgf.13387,7265057,9318552}.
When we capture a scene under room lighting and use the captured image as the target image, radiometric compensation techniques can compute projector pixel colors that replicate the scene in the projected result.
However, most of the previous techniques do not take inter-reflections into account.
While some solutions address the inter-reflection issue~\cite{4392749,Siegl2017}, they are not scalable for systems with many projectors since they require the computation of the inverse of a large matrix (the light transport matrix), the dimensions of which increase linearly with the number of projectors.
In contrast, our approach proposes an illuminant reproduction technique in a distributed manner, where no inversion computation of a large matrix is required and input pixel values for each projector are calculated without relying on information about those for other projectors.
Additionally, the radiometric compensation techniques are based on pixel-wise computation.
Because typical environmental lighting does not exhibit rapid spatial changes in illuminance and chrominance, we compute identical pixel values for a group of projector pixels---and, in some cases, for the entire set of pixels in a projector---to enhance computational efficiency.

Shadows must be eliminated in PM. Researchers have extensively explored this topic using multi-projection systems~\cite{Nagase2011,990943,1272728,7164338} and specialized optics~\cite{10049693}.
However, achieving complete shadowlessness is not our primary goal, as we aim to replicate typical luminaires or diffuse lighting, which naturally results in soft shadows, using projectors.
One solution is to implement light field illumination, which can function as near-diffuse lighting, creating pseudo soft shadows while allowing spatial control of illuminance and chromaticity on scene surfaces~\cite{7223335,10.1145/2992154.2992188,10.1145/3476124.3488624,9664531}.
Nevertheless, light field illumination requires time-consuming calibration to determine the position and orientation of each ray in 3D space.
In contrast, we propose a simpler yet equally effective solution.
Specifically, we employ a large-aperture projector as a diffuse light source to create genuine soft shadows, illuminating the area surrounding the PM target where observers' shadows commonly occur.

\section{Method}
\label{sec:method}

We distribute projectors under the ceiling of the room, among which a portion of the projectors is used for performing PM onto a target object, while the others are used to replicate the environmental lighting provided by the original room lights.
We refer to the former projectors as the \emph{texture projectors} and the latter projectors as \emph{luminaire projectors}.
We establish the correspondence between a point on the scene surfaces and the incident projector pixel by projecting graycode patterns from each projector and capturing them with a wide-angle camera.
To mitigate the seams caused by overlapping projected areas from different projectors, we employ a standard feathering technique~\cite{Bimber2005}.
The remainder of this section elaborates on our two primary contributions.

\subsection{Distributed optimization for reproducing environmental lighting with projectors}
\label{subsec:lighting_reproduction_by_color_patches}

We compute pixel values for the luminaire projectors to accurately replicate the illuminance and chromaticity of environmental lighting in our system, while accounting for inter-reflections among the scene surfaces.
We position the projectors so that the projected areas of different projectors do not significantly overlap, while also avoiding the occurrence of seams between them.
We position color charts on the scene surfaces, following a guideline of placing one or two charts within each projector's frustum.
In cases with significant variations in surface height on the floor or walls, we add additional color charts as needed.
We divide the projected area of each luminaire projector into multiple segments, with each segment covering approximately one of the color charts.
We treat each luminaire projector as a collection of multiple projector nodes, where the projected area of each node corresponds to one of these segments.
Each of the projector nodes for all the luminaire projectors is labeled as $n$.

We input a single pixel value $x_n\in [0,1]$ into each projector node, resulting in the projection of a uniformly colored image from the node.
This approach allows us to sparsely compute optimal pixel values, leveraging the spatially low-frequency characteristics of environmental lighting.
Compared to conventional pixel-wise optimization frameworks~\cite{4392749,Siegl2017}, our low-frequency approach not only reduces computational costs but also mitigates high-frequency artifacts caused by noise in the capturing process.
To ensure scalability with respect to the number of luminaire projectors, we have developed a distributed optimization framework.

\begin{figure}[tb]
  \centering
  \includegraphics[width=\columnwidth]{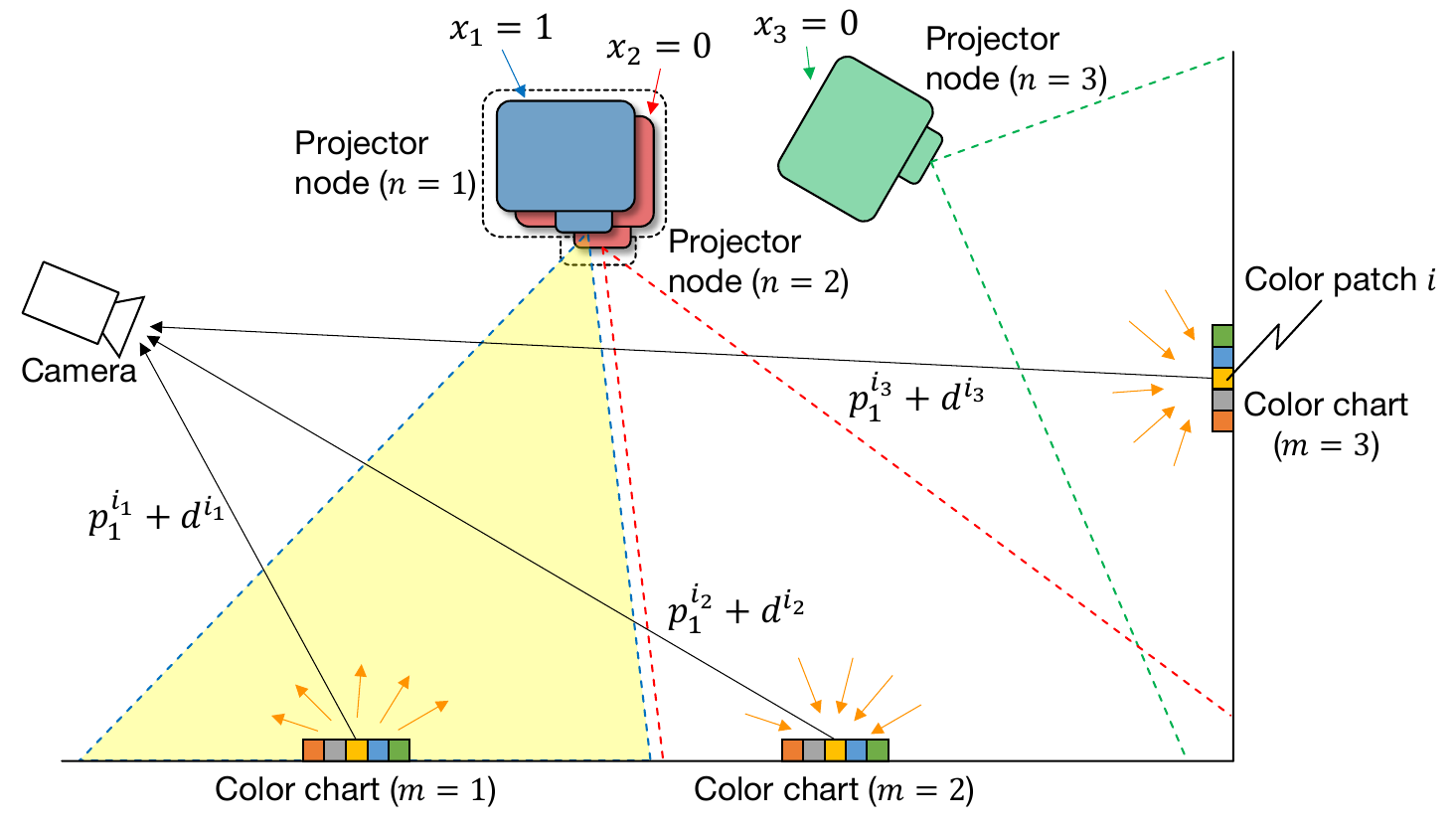}
  \caption{Obtaining attenuation factor $p_1^{i_m}$ by projecting a uniformly white image from projector node 1, while other nodes project uniformly black image. The projected area of a physical projector, depicted by dashed lines, is divided into two segments. Projector nodes 1 and 2 represent virtual projectors, and their projected areas correspond to the respective segments.}
  \label{fig:color_calib_i=1}
\end{figure}

A color chart comprises multiple patches of varying colors and gray levels.
Suppose the reflected RGB color of a patch $i$ in a chart $m$ under the environmental lighting to be reproduced is denoted as $r^{i_m,c}\in[0,1]$, where $c$ represents the color channel (i.e., R, G, or B).
Our goal is to reproduce $r^{i_m,c}$ using the luminaire projectors.
In the following explanation, we omit $c$ for simplicity since the optimization process is color channel-independent.
After deactivating the environmental lighting, we measure a black offset $d^{i_m}$ at each patch by capturing its color under projection of a uniformly black image from all the projectors (i.e., $x_n=0$ for all $n$).
Next, we project a uniformly white image from a projector node $n$ (i.e., $x_n=1$), while projecting a uniformly black image from the other projector nodes (i.e., $x_{n'}=0$ for $n'\ne n$).
After capturing the reflected color of each patch $i_m$, we subtract the black offset $d^{i_m}$ from the captured value, obtaining $p_n^{i_m}\in[0,1]$, which is an attenuation factor representing the proportion of light from a projector node $n$ that is reflected at a color patch $i_m$ (\autoref{fig:color_calib_i=1}).
Therefore, in cases where the projected area of the projector node $n$ does not cover the color chart $m$, $p_n^{i_m}$ represents the effect of inter-reflection.
We repeat this process by changing the projector node that projects a uniformly white image.

The captured color $y^{i_m}$ of each patch under arbitrary illumination by the luminaire projectors can be represented as follows:
\begin{equation}\label{eq:forward_model}
    y^{i_m}=\sum_np_n^{i_m}x_n +d^{i_m}.
\end{equation}
To accurately reproduce the environmental lighting, we minimize the objective function
\begin{equation}\label{eq:objective}
    G=\frac{1}{2}\sum_m\sum_{i_m}\{e^{i_m}\}^2,
\end{equation}
which is the sum of the squared error
\begin{equation}
    e^{i_m}=r^{i_m}-y^{i_m}=r^{i_m}-\sum_np_n^{i_m}x_n-d^{i_m}.
\end{equation}

The gradient of \autoref{eq:objective} is given by
\begin{equation}
    \frac{\partial G}{\partial x_n}=-\sum_m\sum_{i_m} p_n^{i_m}e^{i_m}.
\end{equation}
Therefore, from the difference between the target color and the captured color at a color patch $i_m$ (i.e., $e^{i_m}$), we can compute the gradient for each projector node without relying on information about the other projector nodes.
We update the input pixel values for projector nodes as follows:
\begin{equation}
  x_n[t+1]=\mathcal{P}\left[\frac{1}{K}\sum_{i_m}\left\{x_n[t]-\epsilon\frac{\partial G}{\partial x_n}[t]\right\}\right],
\end{equation}
\begin{equation}
  \mathcal{P}[\bar{x}]:=
    \begin{cases}
        {\bar{x},\quad0\leq\bar{x}\leq1,}\\
        {1,\quad\bar{x}>1,\quad\quad\quad\bar{x}\in\mathbb{R}}\\
        {0,\quad\bar{x}<0,}
    \end{cases}
\end{equation}
where $t$ represents the frame number (or time), and $K$ is the number of color patches in each color chart.
This algorithm is equivalent to a projected gradient method used to minimize $G$, where $\epsilon$ functions as a step size.
It can be proven that this update rule, with sufficiently small $\epsilon$, causes $x_n$ to converge to the optimal value in a suitable sense~\cite{7164338,boyd}.
As described above, we can compute pixel values independently for each projector.
This is a completely distributed process, and furthermore, it is theoretically guaranteed that the addition or removal of projector nodes at runtime does not affect the convergence performance.

\subsection{Large-aperture projector for spatially controlling diffuse lighting on scene surfaces}

\begin{figure}[tb]
  \centering
  \includegraphics[width=\columnwidth]{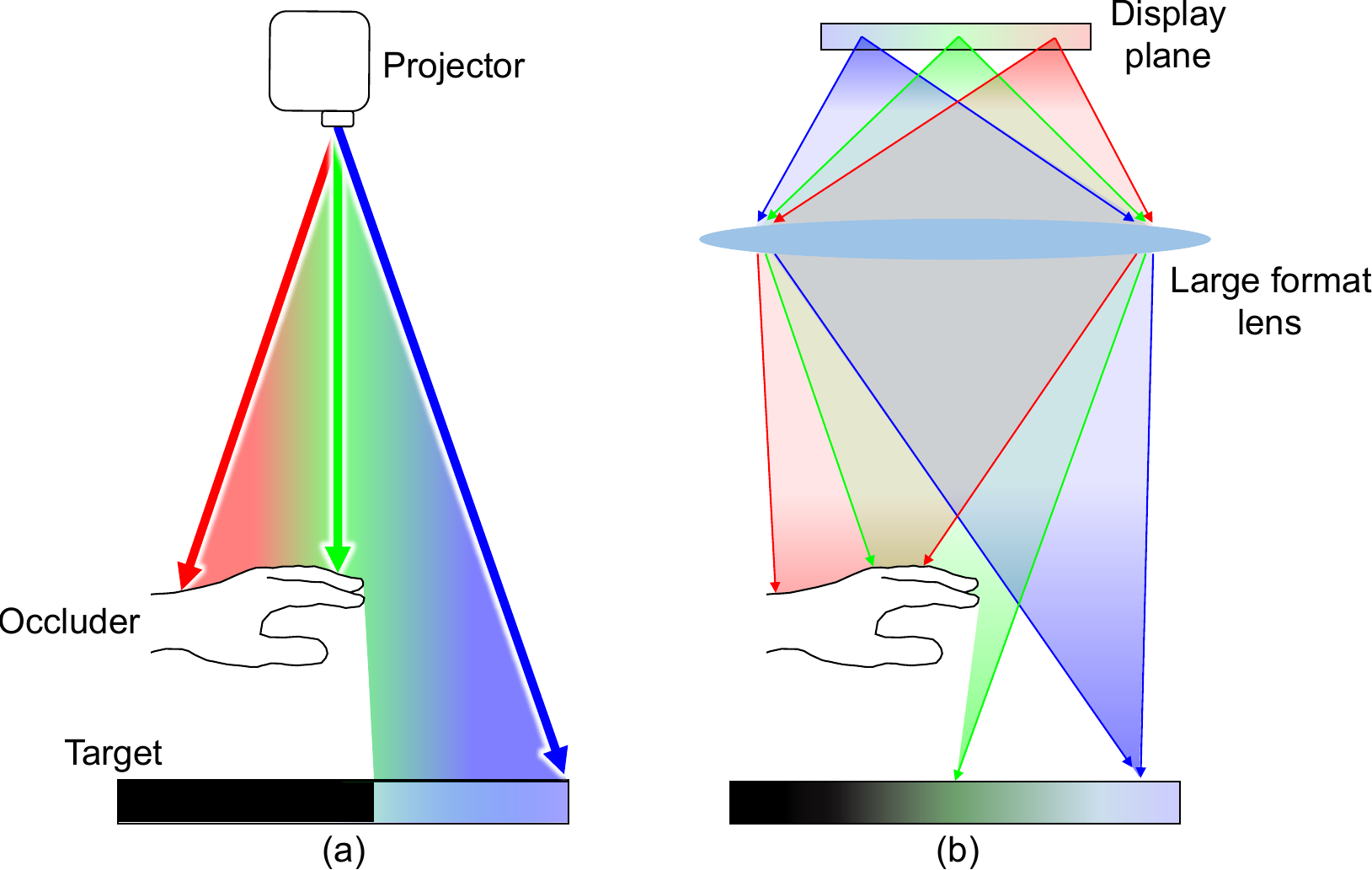}
  \caption{Aperture size of a projector affects the emitted ray intensity and the appearance of a shadow. (a) A standard projector is a point light source producing a high-luminance ray and a hard shadow. (b) The proposed large-aperture projector is an area light source creating a low-luminance ray and a soft shadow.}
  \label{fig:kinds_of_projector}
\end{figure}

\begin{figure}[tb]
  \centering
  \includegraphics[width=\columnwidth]{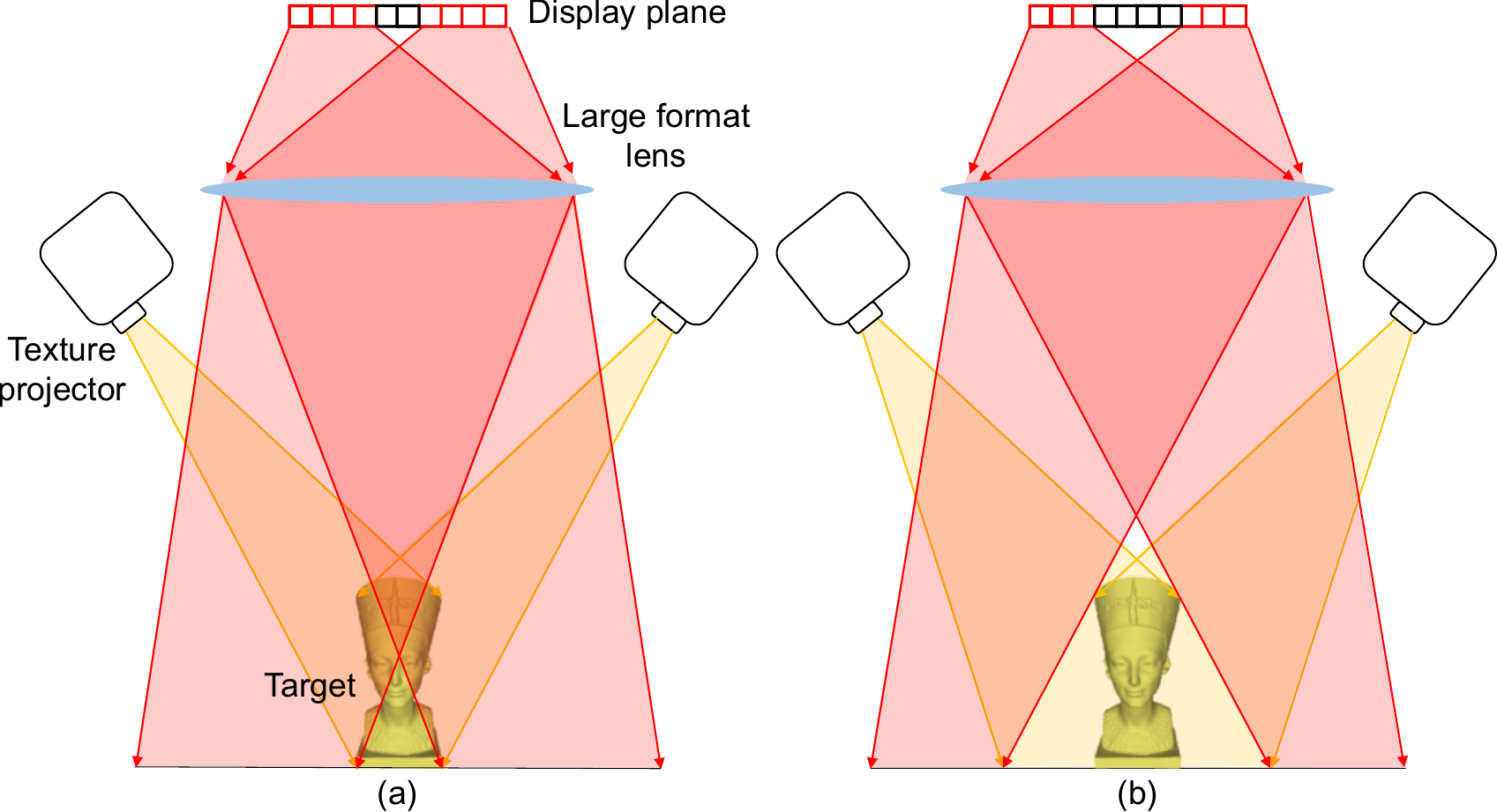}
  \caption{An issue and its solution with the large-aperture projector illuminating the surrounding environmental surfaces of the PM target. (a) Some light rays intended to converge on an environmental surface point near the target unintentionally hit the target surface. (b) Turning off these pixels and compensating for the darkened area with the texture projectors.}
  \label{fig:projected_area_margin}
\end{figure}

Illumination lights incident on the scene surfaces near the PM target tend to be occluded by observers, thus requiring special attention.
When we use a standard projector to illuminate this area, occlusions result in hard shadows (\autoref{fig:kinds_of_projector}(a)), whereas shadows under typical room lighting are soft.
This inconsistency arises from the difference between a point light source that produces directional light (projector) and an area light source that generates diffuse light (typical room light).
To bridge this gap, we propose employing a large-aperture projector as the luminaire for the surfaces near the PM target.
The large-aperture projector comprises a large-format lens and a flat display plane (\autoref{fig:kinds_of_projector}(b)).
When the lens size is sufficiently large, an observer's body, such as a hand or a head, cannot fully occlude all the light from the lens, resulting in soft shadows.
The large-aperture projector offers another advantage compared to a standard projector.
Assuming the illuminance of incident light on a scene surface is identical for both projectors, the luminance of each ray emitted from the lens is significantly lower in the large-aperture projector because the light energy is distributed across the lens.
Therefore, the large-aperture projector provides a more comfortable environment where users experience much less dazzling sensation when they see the projector's lens than under standard projector illumination.

\begin{figure}[tb]
  \centering
  \includegraphics[width=\columnwidth]{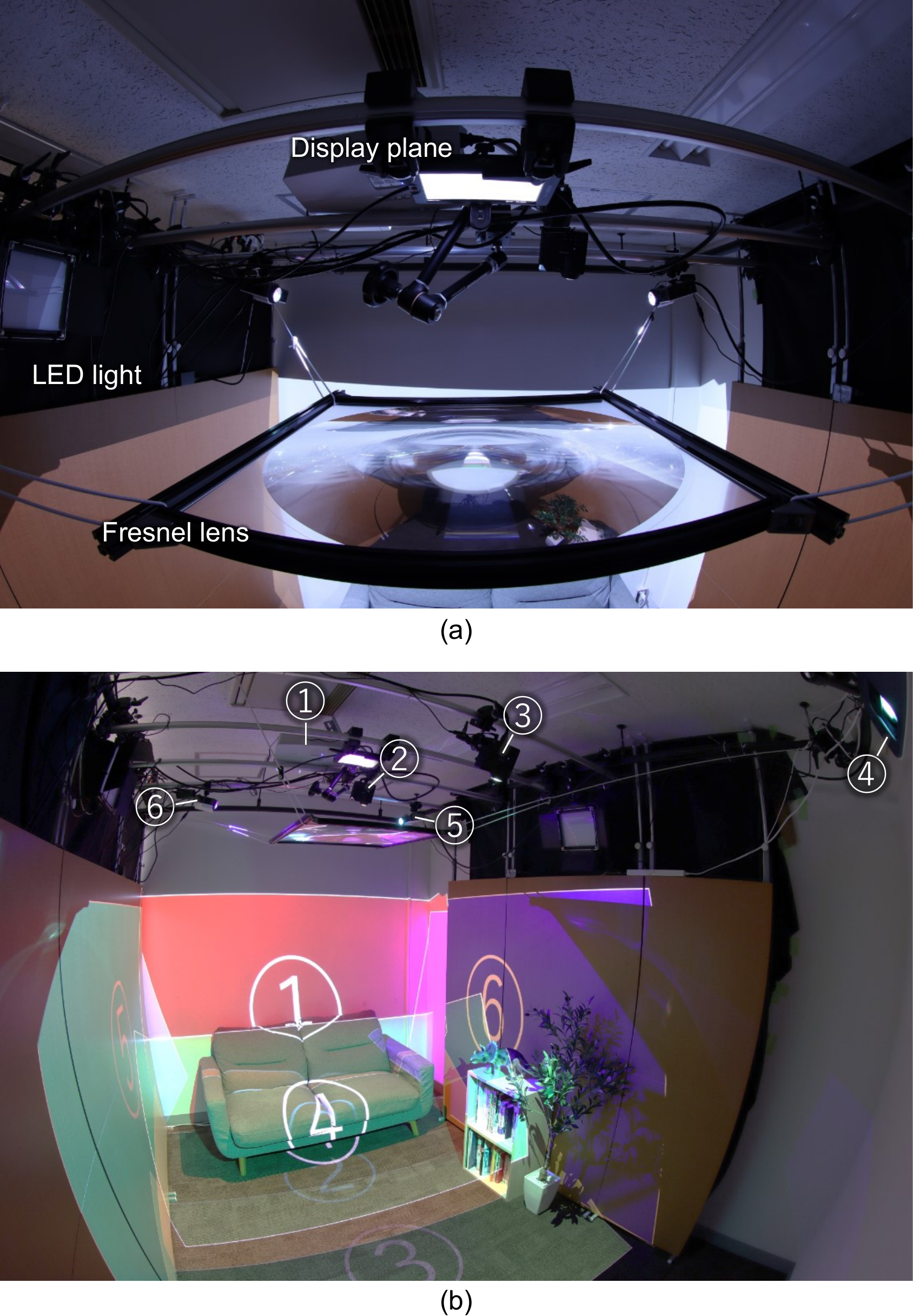}
  \caption{Experimental system. (a) A close-up view of our large-aperture projector. (b) An overview where standard projectors project uniformly colored images.}
  \label{fig:exp1_setup}
\end{figure}

A straightforward method to illuminate the scene surfaces using the large-aperture projector while avoiding the PM target is to turn off the pixels that focus on the target.
However, some light rays emitted from other pixels may still hit the target, as illustrated in~\autoref{fig:projected_area_margin}(a).
This significantly degrades the contrast of the projected results on the target.
We address this issue by simply turning off a larger area of pixels on the display plane of the large-aperture projector, which includes all the pixels emitting light rays that hit the target.
While this solution mitigates contrast degradation, it may cause some of the scene surfaces around the PM target to darken.
To compensate for the reduction in illuminance, we utilize the texture projectors to function as luminaire projectors for the darkened area of the scene surfaces (\autoref{fig:projected_area_margin}(b)).

\section{Experiment}
\label{sec:experiment}

We conducted both quantitative and qualitative experiments including user studies to validate the proposed method.

\subsection{Experimental system}

\autoref{fig:exp1_setup} shows our experimental system.
We arranged a sofa, desk and other furniture in our lab space measuring 2800$\times$2200 mm to recreate a typical living room.
Additionally, we placed small objects on the desk and bookshelf, including a white bust statue of 200 mm height on the desk to serve a projection target.
We measured the reflectance of the object surface using a spectroradiometer (Topcon SR-LEDW) and a reflection standard white ceramics plate (Evers Corporation EVER-WHITE No. 9582), then averaged the measured reflectance values within the visible spectral range. The resulting average value was 0.56.
Our method is effective for other objects with unknown reflectance properties, provided they have a diffuse surface. In such cases, radiometric compensation techniques~\cite{7265057,9318552} are applied to reproduce the desired appearances.
For capturing the scene, we used a camera (Canon EOS M6 Mark II) equipped with a fisheye lens (Canon EF8-15 mm f/4L Fisheye USM) positioned at the edge of the room.

\begin{figure*}[tb]
  \centering
  \includegraphics[width=\linewidth]{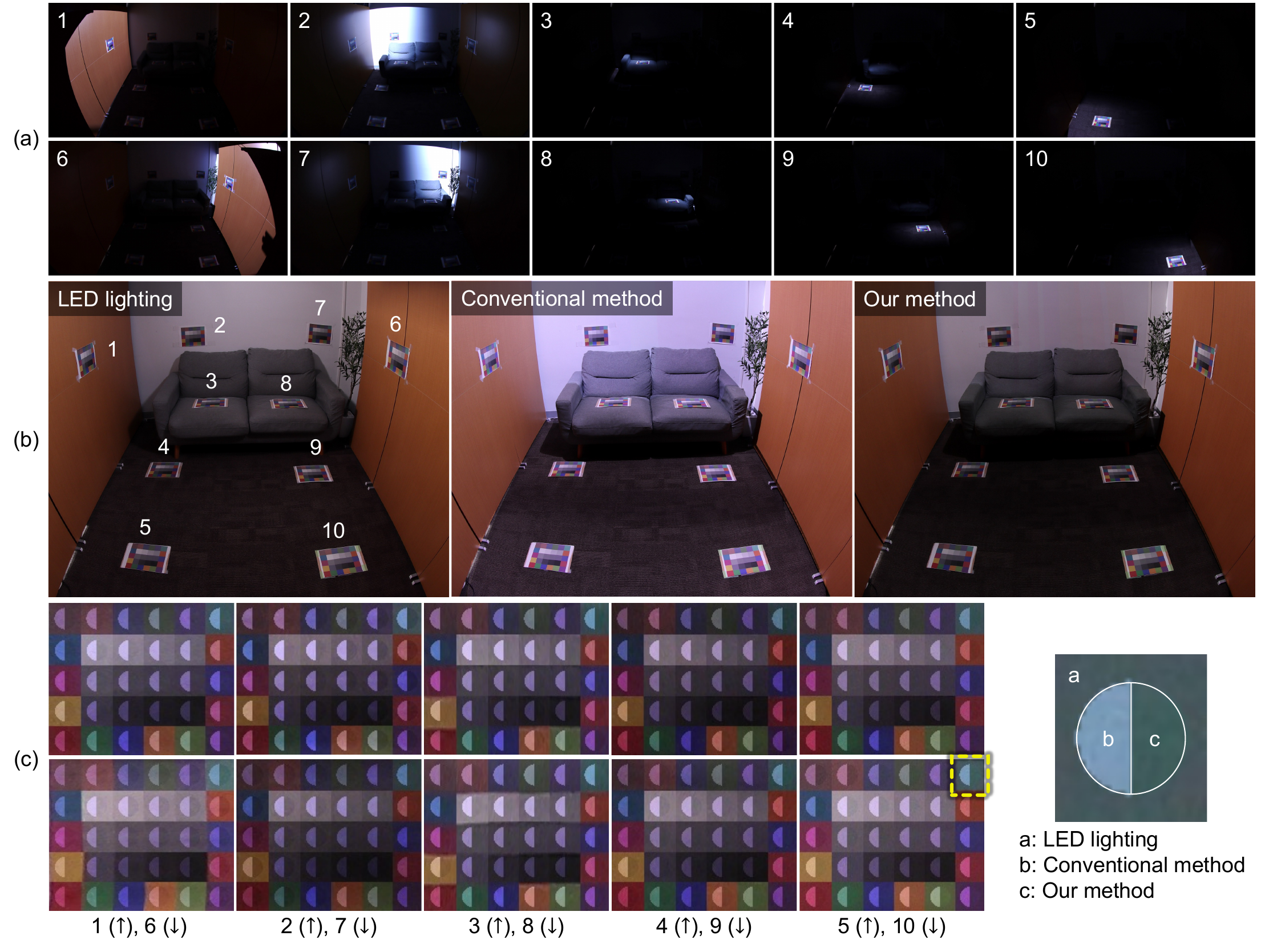}
  \caption{The experimental results of the environmental lighting reproduction. (a) Projecting a uniformly white image from each projector node (indicated by the overlaid number) for obtaining $p_n^{i_m}$. (b) Captured images under three different types of environmental lighting; typical LED lights (target appearance), the reproduced lighting with projectors using a conventional method~\cite{1381255}, and that using our method. (c) The appearance of the color chart under the three lighting conditions.}
  \label{fig:lighting_reproduction_combined}
\end{figure*}

We employed six projectors as luminaire projectors, comprising five standard projectors and one custom-made large-aperture projector.
Among the standard projectors, two (Optoma ML1050ST+S1J) were directed at the side walls, one (Acer H6517ST) was directed at the front wall, and two (RICOH PJ WXC1110) were directed at the floor.
The large-aperture projector, comprising an LCD display (OSEE T7) and a 500$\times$500 mm fresnel lens (SIGMAKOKI FRLN-500S-250P), provided illumination for the desk.
We designated a standard projector (BenQ TK850) as the texture projector for projecting textures onto the target surface.
The projectors and the camera were controlled by a PC (CPU: AMD Ryzen 9 3900 X 12-Core Processor, RAM: 64 GB).
For typical environmental lighting sources, we utilized LED-based luminaires (NEEWER NL660 Bi-Color LED Panel Light) with adjustable luminance and color temperature settings.
All of these projectors and LED lights were mounted on the ceiling.
As the step size of our method, $\epsilon=0.00001$ was used in all the experiments.

\subsection{Reproduction of environmental lighting}\label{subsec:exp_reproduction}

We validated our method of reproducing typical environmental lighting using projectors, explained in \autoref{subsec:lighting_reproduction_by_color_patches}.
We positioned a total of ten color charts on the front and side walls, as well as the floor.
Then, we captured the reflected RGB color of each patch under the LED lighting and stored it as $r^{i_m}$.
Next, after turning off the LED lights, we measured the black offset $d^{i_m}$ at each patch.
We then measured the attenuation factor $p_n^{i_m}$ by projecting a white image from each projector node and capturing the reflected color at each color patch as shown in~\autoref{fig:lighting_reproduction_combined}(a).

The initial input pixel values in the iteration process of our method (i.e., $x_n[1]$) were computed using a conventional radiometric compensation technique~\cite{1381255}.
This technique relies on the same linear projector-camera reflection model as \autoref{eq:forward_model}, but it does not account for inter-reflection.
Specifically, assuming the projected area of the $n$-th projector node covers the color chart $m'$, the conventional technique reformulates \autoref{eq:forward_model} as $y^{i_{m'}}=p^{i_{m'}}_nx_n+d^{i_{m'}}$.
Then, it solves this equation for $x_n$, given the target appearance $r^{i_{m'}}$, yielding $x_n=(r^{i_{m'}}-d^{i_{m'}})/p^{i_{m'}}_n$, which is used as the initial input value in our method.

\begin{figure}[tb]
  \centering
  \includegraphics[width=\columnwidth]{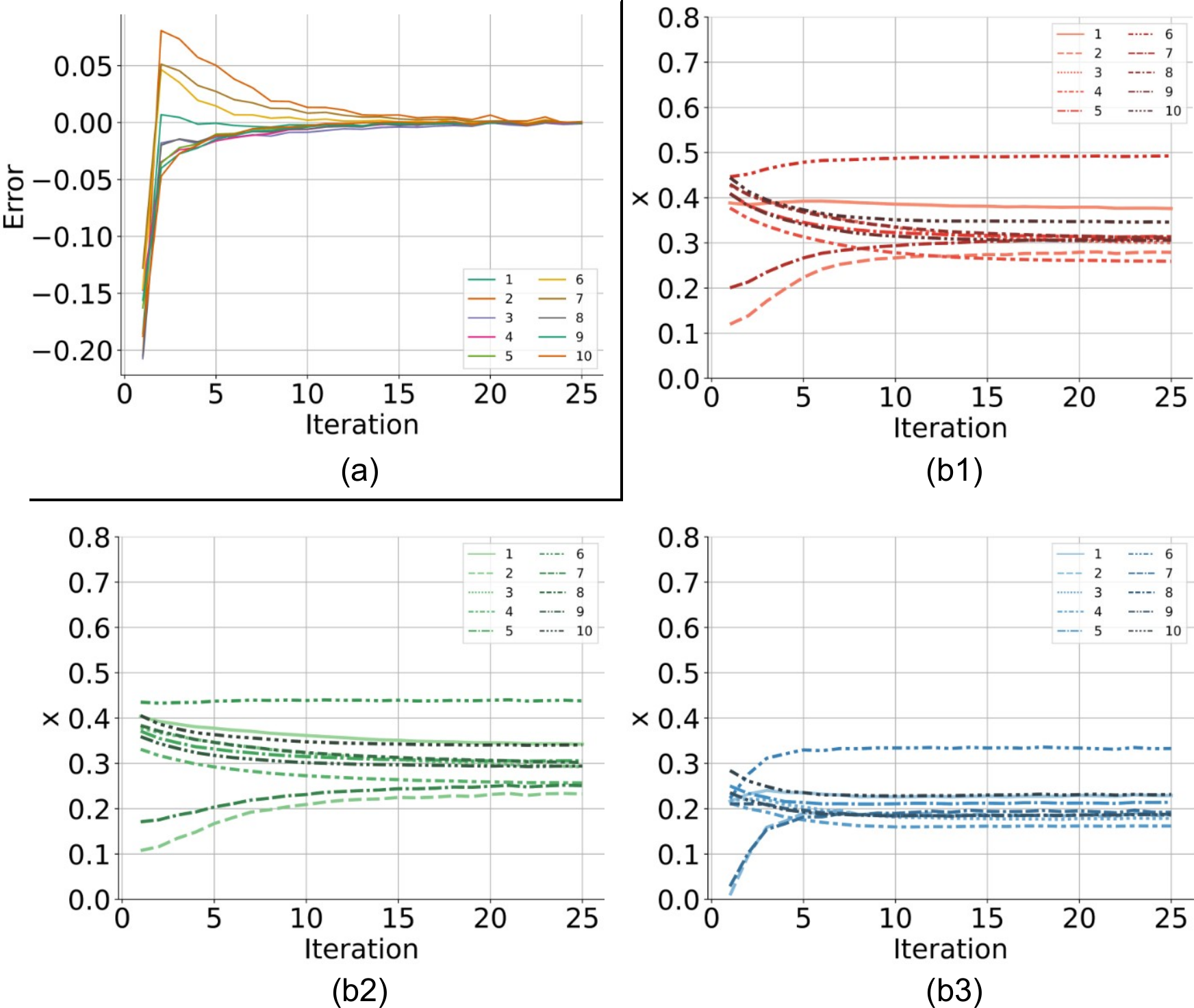}
  \caption{Time series of data in the proposed distributed optimization technique over 25 iterations. (a) The averaged error observed in each color chart. (b) The input pixel values of each color channel for each projector node (b1: R, b2: G, b3: B).}
  \label{fig:error}
\end{figure}

\autoref{fig:lighting_reproduction_combined}(b) shows the appearance of our room under three different lighting conditions: the LED lights (representing the target appearance), the luminaire projectors using the conventional method~\cite{1381255}, and the luminaire projectors using the proposed method after $t= 20$ iterations.
It took 9 seconds to compute the projection images, project them, and capture the projected results in each iteration. Therefore, the entire process of 20 iterations requires about 180 seconds. When focusing on the computation time alone, it took about 18 seconds for 20 iterations.
\autoref{fig:lighting_reproduction_combined}(c) showcases the captured colors of the color charts under these conditions.
These results underscore the ability of our method to more accurately replicate the target environmental lighting compared to the conventional method, particularly emphasizing the significant brightness elevation in the conventional method, which does not account for inter-reflection.
\autoref{fig:error} shows the transition of the output of our objective function (\autoref{eq:objective}) through the iteration.
The figure also includes the transition of the input pixel values for the projector nodes in all three color channels.
These results demonstrates the convergence of our method in a distributed cooperative manner within around 20 iterations.

\begin{figure}[tb]
  \centering
  \includegraphics[width=\linewidth]{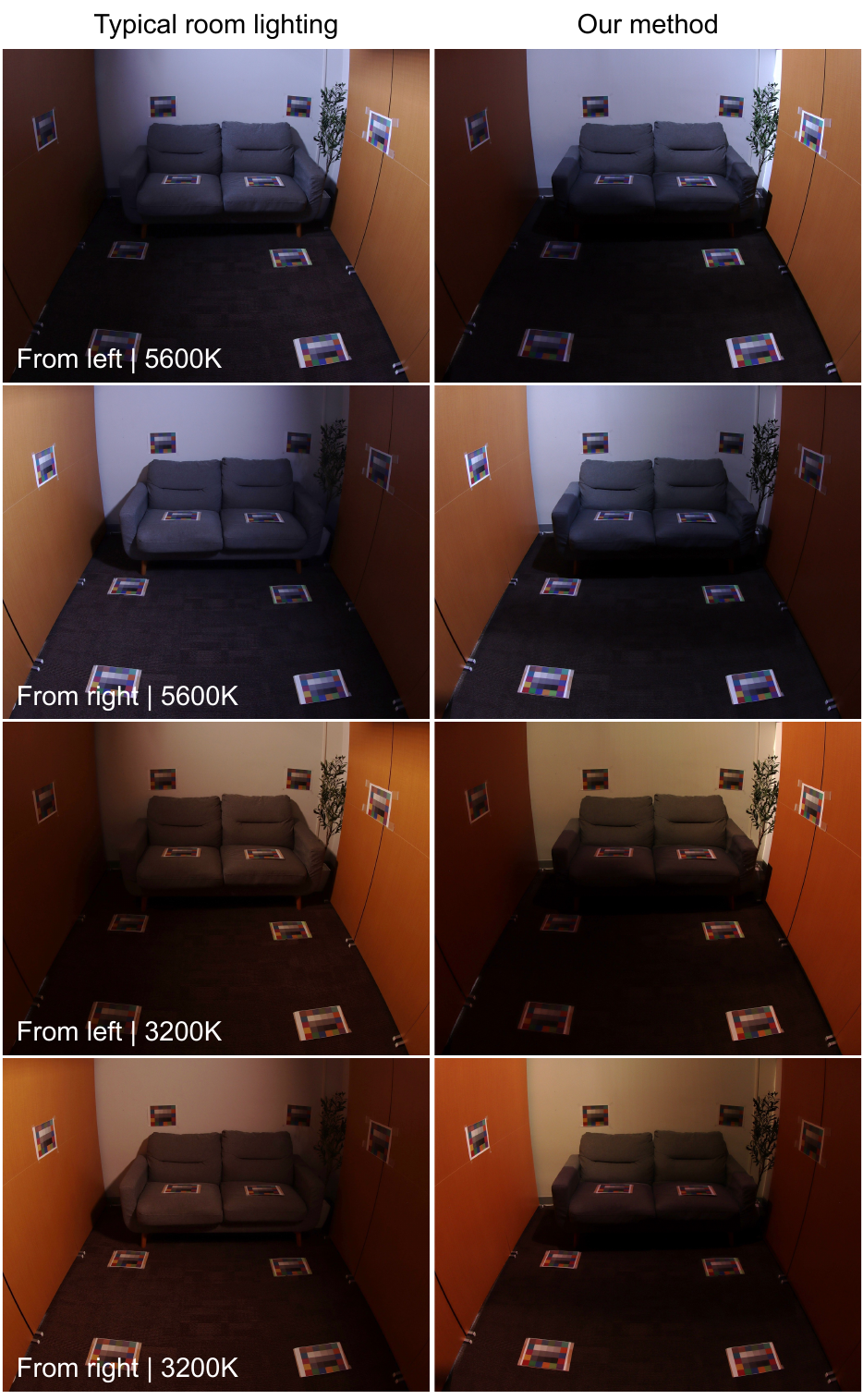}
  \caption{Captured scene appearances (left) under four different environmental lighting conditions with varying light positions and color temperatures, and (right) those under our projector-based environmental lighting.}
  \label{fig:lighting_compare}
\end{figure}

We conducted a qualitative evaluation of our technique by preparing different target environmental lighting conditions.
Specifically, we altered the position and color temperature of the LED light, resulting in four distinct room appearances.
\autoref{fig:lighting_compare} shows these new target appearances, alongside the replication results achieved using our method with the luminaire projectors.
We can confirm that the target appearances were accurately reproduced at a level where we can easily discern how the position and color temperature of the original LED light are altered.

\subsection{Texture manipulation of the projection target}\label{subsec:exp_PM}

\begin{figure}[tb]
  \centering
  \includegraphics[width=\linewidth]{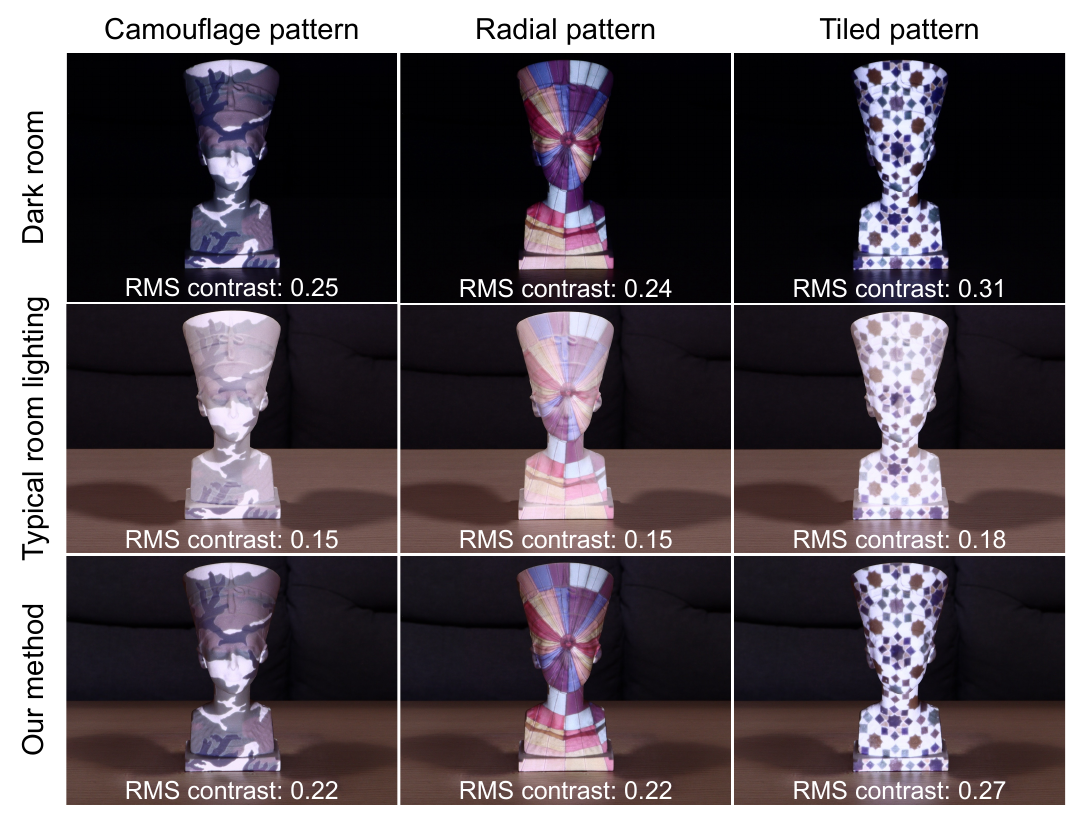}
  \caption{PM results of various texture patterns onto a white bust statue under three different lighting conditions; dark room, typical LED lighting, and our method.}
  \label{fig:texture_results}
\end{figure}

By conducting PM in the prototype, we verified the ability of our method in alleviating the contrast degradation of PM under environmental lighting.
\autoref{fig:teaser} shows the white bust statue positioned on the desk before and during PM of a crack pattern under three lighting conditions: dark room, typical environmental lighting by LED lights, and the proposed replicated environmental lighting.
The images captured before PM, as shown in~\autoref{fig:teaser}(b), demonstrate that our method can selectively remove environmental light from the target object, while typical lighting significantly increases the brightness of the target.
The PM results shown in~\autoref{fig:teaser}(c) demonstrate that our method can manipulate the target appearance under environmental lighting while minimizing the intensity elevation in dark areas of the projected texture compared to the typical lighting.
A similar effect of alleviating the contrast degradation is also exhibited in the PM results of other textures in~\autoref{fig:texture_results}.

\autoref{fig:teaser} and \autoref{fig:texture_results} also include the root mean square (RMS) contrast values of the PM results.
The dark room condition yielded the highest contrast among all PM results, with the typical environmental lighting producing the lowest contrast.
We quantitatively confirmed that the proposed method yielded a higher level of contrast compared to typical lighting.
We also measured ANSI contrast by replacing the bust statue with a planar surface (200$\times$200 mm) on the desk and projecting a 4$\times$4 checker pattern onto the plane using the texture projector.
We measured the luminance values of each checker area using the spectroradiometer.
The ANSI contrast values were 180:1, 30:1, and 65:1 in the dark room condition, the typical environmental lighting condition, and the proposed replicated environmental lighting condition, respectively.
We confirmed that the proposed method produced a contrast ratio more than two times higher than that obtained under the typical lighting condition.

We altered the surface material of the target object to a mirror-like specular material using PM under the same three lighting conditions.
To simulate the glossy appearance of the object in the real environment from an observer’s viewpoint, we generated the projection image using a conventional 2-pass rendering technique~\cite{Bimber2005} as follows.
As preparation, we measured the color and 3D shape of the experimental room using an RGB-D camera (Apple iPad Pro 5th gen), calibrated the projector's intrinsic and extrinsic parameters, and measured the position of the observer's eyes (or the camera capturing the experimental result) in the room in advance.
First, employing a game engine, specifically Unity, we reproduced the experimental room in a virtual space and placed the 3D model of the bust statue in the same location as the physical setup.
Then, we rendered the bust statue with a specular material and captured it with a virtual camera placed at the observer's location, constituting the first rendering pass.
The captured image was then mapped onto the target object from the virtual camera position using the projective texture mapping technique.
In the second rendering pass, the mapped object was captured by another virtual camera with intrinsics and extrinsics identical to the projector's.
Finally, the captured image was used as the projection image for the physical space.

\autoref{fig:ex_specular} displays the PM results.
In the dark room condition, the PM result exhibits the highest contrast.
However perceiving its surface material as specular is challenging because it is not physically correct for specular surfaces to appear visible when the surrounding environment is not visible.
The PM result in the typical environmental lighting condition shows the lowest contrast due to the increased black offset caused by the lighting, which degrades the realism of the projected material.
In contrast, the PM result in the proposed method exhibits higher contrast than the typical lighting condition, and the surrounding environment remains visible.
As a result, the projected bust statue can be perceived as having a specular material to the greatest extent under the proposed condition.

\subsection{User study 1: Perceived color mode of projected results}\label{subsec:exp_userstudy1}

We conducted a user study to investigate how much the proposed method affects the observer's perceived color mode.
In vision science, it has been well known that there are several perceptual modes of color appearance, with surface-color and aperture-color modes being the most common in our everyday lives~\cite{Katz1935,Uchikawa2001}.
These modes are determined by the luminance relationship between an observed object and its surroundings~\cite{Uchikawa1989}.
The proposed technique replicates the environmental lighting, resulting in bright surroundings.
Therefore, assuming observers perceive PM results in a dark room in aperture-color mode, the hypothesis in this study is:
\begin{description}
    \item[H1] The perceived color mode for PM results is shifted toward the surface-color mode by the proposed environmental lighting.
\end{description}

We showed projected results to each participant in two lighting conditions, the dark room condition and the proposed replicated environmental lighting condition, and asked them to rate the perceived mode of its appearance.
We excluded the typical environmental lighting condition in this study because the appearance of a PM result is significantly affected by the room lights, and at times, the details of the projected content become invisible.
We used the bust statue as the projection target and presented the PM results of four projection images, including the crack pattern (\autoref{fig:teaser}), as well as the camouflage pattern, radial pattern, and the tiled pattern as shown in~\autoref{fig:texture_results}.
Thus, each participant completed eight trials in total (2 lighting conditions$\times$4 projection images).
In each trial, the participants sat 1.5 m away from the projected object and adapted to the surround environment for 5 min.
After that, they looked at the projected object and answered a five-point scale question about the perceived mode (1: aperture-color mode, $\ldots$ , 5: surface-color mode).
There was no time limit for the participant to look at each stimulus before deciding the response.
The study was approved by the Research Ethics Review Committee of the institute to which the corresponding author belongs (approval number; R5-6).

\begin{figure}[tb]
  \centering
  \includegraphics[width=\columnwidth]{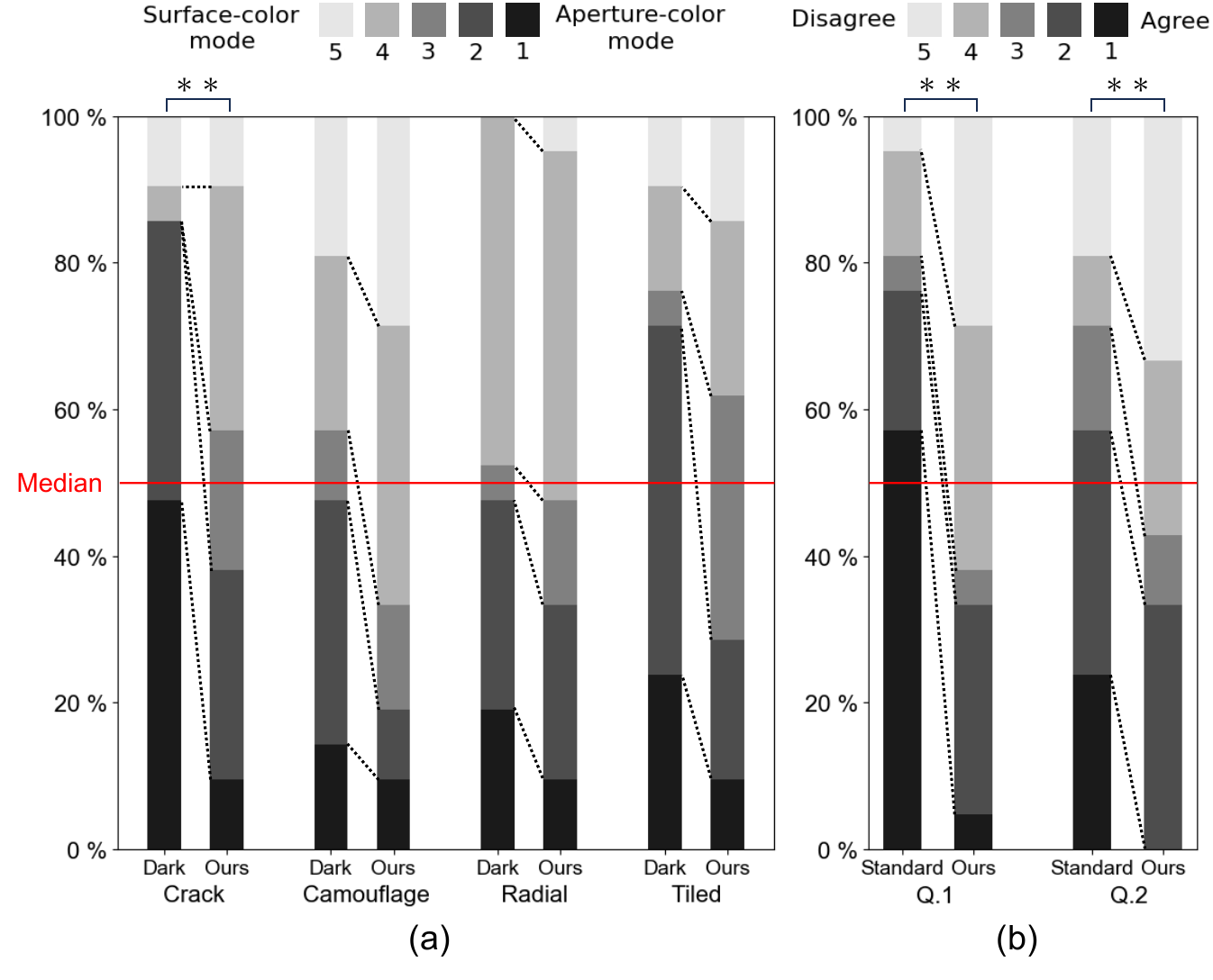}
  \caption{Summary of user study results: (a) the first study described in~\autoref{subsec:exp_userstudy1}, and (b) the second study described in~\autoref{subsec:exp_userstudy2}.}
  \label{fig:user_study_result}
\end{figure}

Twenty-one participants (20 males and 1 female, aged 21 to 27 years old, university students) volunteered for the study.
All participants were na\"{i}ve to the purpose of the experiment, had corrected visual acuity, and gave informed consent.
Before the experiment, we instructed participants that the aperture-color mode and surface-color mode respectively correspond to the qualities of color appearance for a self-luminous object and an illuminated surface, respectively.
All participants agreed that they understood the concept of these color modes.
The order of experimental conditions was randomized and balanced across participants.
\autoref{fig:user_study_result}(a) summarizes the participants' responses.
The median scores were shifted towards the surface color modes by the proposed environmental lighting for all the projected patterns.
A Wilcoxon signed rank test showed that there was a significant difference (p$<$0.01) between scores given for the proposed replicated environmental lighting compared to the dark room condition when the crack pattern was projected.
On the other hand, the test did not reveal any significant differences for the other projected patterns (p$\geq$0.05).
Therefore, the hypothesis of H1 was partially supported.
The results indicate that our environmental lighting replication method provides a better PM environment for surface material editing with reduced self-luminous appearance, making it a more natural choice than a typical dark room.
We will discuss the results in more details in~\autoref{sec:discussion}.

\subsection{User study 2: Subjective evaluation of large-aperture projector}\label{subsec:exp_userstudy2}

\begin{figure}[tb]
  \centering
  \includegraphics[width=\columnwidth]{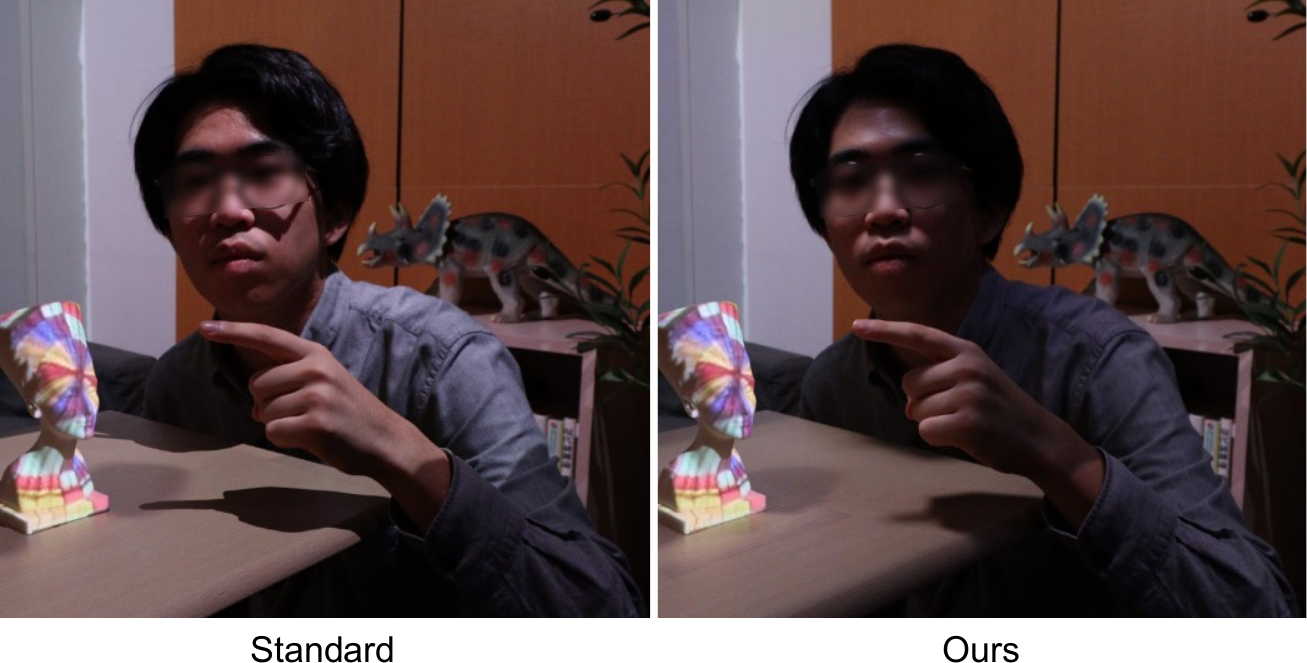}
  \caption{Comparison of shadows created by a standard projector and our large-aperture projector. Note that the eyes are intentionally blurred to protect the individual's identity.}
  \label{fig:compare_normal_and_ours}
\end{figure}

\begin{figure*}[tb]
  \centering
  \includegraphics[width=\linewidth]{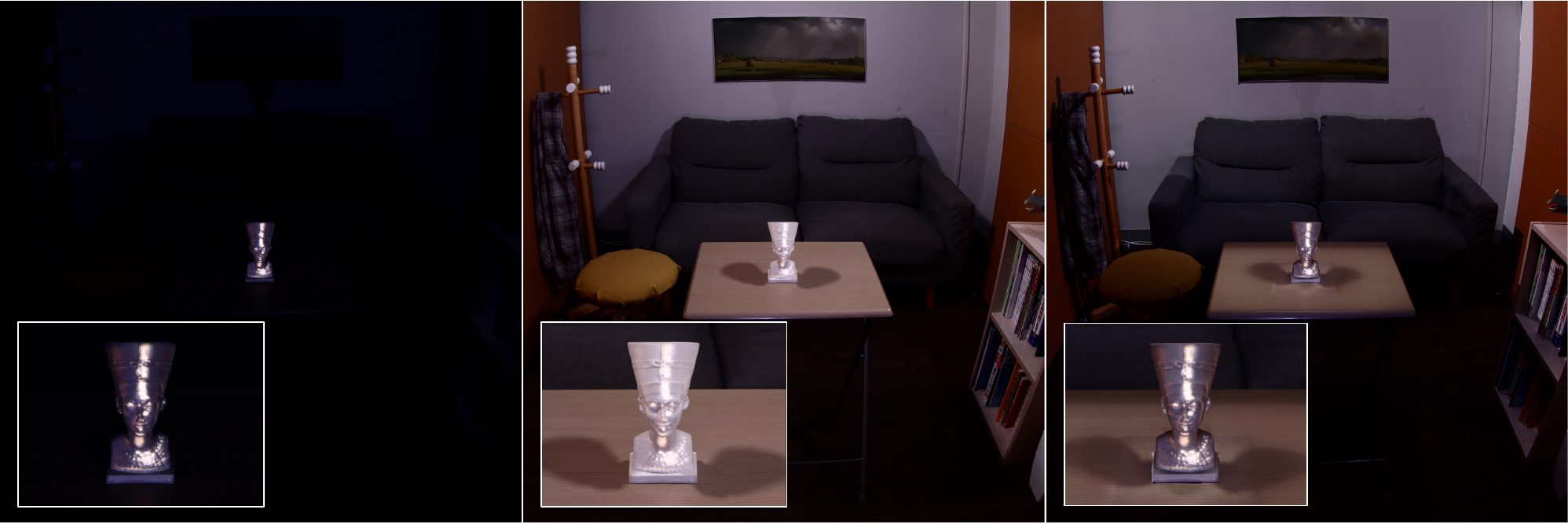}
  \caption{Visually altering the surface material of the diffuse statue to a mirror-like specular appearance in three lighting conditions: (left) in a dark room, (center) under typical room lighting, and (right) under the proposed projector-based environmental lighting.}
  \label{fig:ex_specular}
\end{figure*}

In our second user study, we evaluated our large-aperture projector as a luminaire for illuminating the surrounding environmental surfaces of the projection target, comparing it to a standard off-the-shelf projector.
We focused on the following two issues:
First, the illumination produced by the standard projector appears to result from directional light, which is inconsistent with common diffuse environmental lighting.
In contrast, the illumination produced by the proposed large-aperture projector seems to reduce the perception of directional light.
Secondly, the standard projector, considered a point light source, emits high-luminance rays and creates hard shadows, while the proposed projector, seen as an area light source, generates much lower-luminance rays and softer shadows.
Hard shadows appear, disappear, and change shape significantly even with slight movements of an occluder.
In contrast, soft shadows are more temporally stable.
\autoref{fig:compare_normal_and_ours} compares the shadows created by a standard projector with those created by the proposed large-aperture projector.
Regarding the emitted rays, observers may experience a dazzling sensation only when the standard projector is in use and they possibly see the projector lens.
Consequently, the illumination produced by the standard projector has the potential to disturb users' tasks in PM due to the presence of hard shadows and high-luminance emitted rays, in contrast to the proposed large-aperture projector.
To investigate these aspects, this study specifically set the task as observers discussing the surface material of a PM target with each other and examined the following hypotheses.
\begin{description}
    \item [H2] The proposed projector mitigates the impression of directional light compared to a standard projector.
    \item [H3] The proposed projector alleviates the optical disturbance in communication compared to a standard projector.
\end{description}

While one might think that H2 is evident because the impression of directional light primarily depends on how much an observer pays attention to the shadows, we believe H2 addresses an important aspect. That is because even a small bump causes a hard shadow in a standard projector condition. Given that such small bumps are common on real-world surfaces, users would unconsciously notice them without explicitly focusing on the shadows.

Each participant and an experimenter sat by the projection target and engaged in a conversation about the material impression of the PM results, while the experimenter changed the projected textures among those introduced in this paper so far including the mirror-like specular texture according to the conversation.
Note that they did not discuss the environmental lighting.
The conversation lasted for about 5 minutes with the first 2.5 minutes and the rest conducted under different projector conditions.
The participant was allowed to use their hands in the conversation, and the experimenter frequently pointed to the projection object to create shadows.
Afterward, the participant was asked two questions and answered on a five-point scale questionnaire (1: agree, $\ldots$ , 5: disagree).
The questions were: (1) Was the environmental lighting directional? and (2) Did the environmental lighting and shadows disturb the communication?
The study was approved by the Research Ethics Review Committee of the institute to which the corresponding author belongs (approval number; R5-6).

The same participants from user study 1 volunteered for this study.
All participants were na\"{i}ve to the purpose of the experiment.
The order of experimental conditions was randomized and balanced across participants.
\autoref{fig:user_study_result}(b) summarizes the participants' responses.
A Wilcoxon signed-rank test showed significant differences (p$<$0.01) between the scores given for the proposed projector condition compared to the standard projector condition in both questions.
Specifically, the environmental lighting provided by the standard projector was perceived as significantly more directional than that provided by the proposed large-aperture projector.
Additionally, the participants judged that the illumination by the standard projector disturbed the communication significantly more than that by the proposed projector.
Therefore, the hypotheses of H2 and H3 were both supported.
We confirm that the proposed large-aperture projector provides more consistent illumination with typical environmental lighting and offers a better environment for communication about the projected objects and potentially for other tasks.

\section{Discussion}
\label{sec:discussion}

Through a series of quantitative and qualitative experiments, we demonstrated the effectiveness of our method for achieving PM in a well-lit room.
The experiment in~\autoref{subsec:exp_reproduction} showed that typical environmental lighting is reproducible to the extent that we can recognize the locations and color temperatures of the original light sources.
The remaining discrepancies between the target and reproduced appearances can be mitigated by placing the color charts more densely in the scene.
This adjustment does not require any modifications to our framework but increases the computational cost.
The accuracy of the reproduction depends on the application, and we believe that there are not many applications requiring very accurate reproduction because PM users typically do not focus on the environment.

The experiment in~\autoref{subsec:exp_PM} demonstrated that our replicated environmental lighting mitigates contrast degradation in PM results compared to typical environmental lighting.
As shown in \autoref{fig:ex_specular}, our framework effectively reproduces the reflection of light from a scene surface onto the PM target in a physically accurate manner, by which the realism of the specular reflection was significantly improved.
We can naturally extend this framework to replicate bidirectional light field interactions between the PM target and its surroundings.
This study represents a promising first step toward achieving \emph{perceptual realism} in the context of PM, defined as the production of imagery indistinguishable from real-world 3D scenes~\cite{10.1145/3478513.3480513}.

The first user study in~\autoref{subsec:exp_userstudy1} confirmed a tendency for observers to perceive the PM results in surface-color mode when using the proposed environmental lighting.
However, the analysis did not show this tendency to be statistically significant in three out of four projected textures.
A possible explanation for this result is as follows: Previous psychophysical experiments used visual stimuli consisting solely of color patches or spatial patterns without any contextual information~\cite{Uchikawa1989,Uchikawa2001,10.1167/jov.21.13.3}.
In contrast, our stimuli provided ample contextual information for participants to identify the object as a bust statue and understand the meaning of the projected images.
Additionally, all participants knew that the appearance of the target object was altered in PM, suggesting that they might have implicitly understood that the object reflected light from the projector rather than being self-luminous.
These factors may have biased their judgments.
On the other hand, it is worth mentioning that even under such conditions, there was a case where the proposed method caused a significant shift from aperture-color mode to surface-color mode.
To the best of our knowledge, we are the first to focus on the perceived color mode in PM.
We believe that our research is essential because we investigated the color mode in real PM scenarios and demonstrated that environmental lighting influences the perceived color mode, albeit with the degree of influence varying depending on the projected contents.

Through the second user study (\autoref{subsec:exp_userstudy2}), we confirmed that selecting a large-aperture projector as the luminaire for illuminating the surrounding environmental surfaces of the projection target is crucial to creating a lighting environment familiar to users and minimizing disruptions during their tasks.
Specifically, the large-aperture projector significantly reduced the impression of directional light which is uncommon in typical environmental lighting, and substantially decreased bothersome high-luminance emitted rays and hard shadows during communication compared to a standard projector.
Additionally, we found that the standard projector produces intense (potentially dazzling) specular reflections, whereas the proposed projector generates much weaker specular reflections.
Therefore, an important design guideline that we have learned from this result, when considering the replacement of room lights with projectors, is the necessity of incorporating a projection system that, alongside standard projectors, can function as an area light source to illuminate areas in proximity to the projection target.
Such projection systems include multi-projection systems, light field projectors, and large-aperture projectors.
An interesting avenue for future work would be to compare the performance of these different types of area light projectors in this context.
On the other hand, we applied a standard projector to the texture projector in the current setup, resulting in a hard shadow on the PM target when a user approached it. We consider applying area light projectors as texture projectors to be another important area for future work.

Although our evaluated setup mimics a living room, it is primarily designed for special cases of PM. The camera used for calibration must be stable in the room and observe example lighting before installing the projectors. Here, we explain why we chose a living room for PM. We envision a future where more projectors replace standard luminaires in ordinary rooms, and PM supports our daily activities. With this vision, we selected a living room as a testbed to address the typical dark environment constraint, which becomes more significant when PM is applied in ordinary rooms. While the necessity of a stably fixed camera for distributed projector optimization is a current limitation, we could potentially relax this limitation by representing the entire PM system using a neural network~\cite{10269043}. Once the neural representation is established by casually capturing the projected scene using a hand-held camera, we can optimize the projected images to reproduce various lighting conditions without physically projecting calibration images during the optimization process. Another limitation is the necessity of capturing example lighting. We believe that an interesting avenue for future work would involve using professional illumination simulation software applied in the illumination design industry, such as DIALux\footnote{\url{https://www.dialux.com/}}, for synthesizing example lighting rather than capturing it.

\section{Conclusion}
\label{sec:conclusion}

In this paper, we discussed and validated a technical solution for achieving PM in environmental lighting.
As an initial step toward this intriguing goal, we attempted to replace room lights with projectors to selectively illuminate the scene, excluding the projection target.
We proposed two technical contributions in this paper: a distributed projector optimization framework designed to replicate environmental lighting and the application of a large-aperture projector as an area light source to illuminate areas near the projection target.
Thanks to these contributions, we achieved two significant outcomes: (1) accurate replication of environmental lighting using projectors and (2) the mitigation of disturbances to users' tasks in PM by reducing undesirable high-luminance emitted rays and hard shadows that typically occur with standard projectors.
A PM experiment demonstrated that our projector-based luminaire could enhance the contrast in PM results under environmental lighting compared to typical lights, thereby increasing the realism of projected results.
Additionally, we confirmed that it allows for a shift in perceived color mode from the undesirable aperture-color mode, where observers perceive the projected object as self-luminous, to the surface-color mode in PM.
As part of our future work, we plan to extend this system to support dynamic PM.

\acknowledgments{%
This work was supported by JSPS KAKENHI grant number JP20H05958, JST PRESTO Grant Number JPMJPR19J2, and the Future Social Value Co-Creation Project, Osaka University.
}

\bibliographystyle{abbrv-doi-hyperref}

\bibliography{export}

\end{document}